\documentclass[prd,showpacs,nofootinbib,preprintnumbers]{revtex4}

\usepackage{graphicx}
\usepackage{amssymb}
\usepackage{pifont}

\def\bfnabla{\mbox{\boldmath $\nabla$}}

\newcommand{\be}{
\begin{equation}
}{\bf }
\newcommand{\ee}{
\end{equation}
}
\newcommand{\bea}{
\begin{eqnarray}
}
\newcommand{\eea}{
\end{eqnarray}
}
\def\als{\alpha_{\rm s}}
\def\siml{{\ \lower-1.2pt\vbox{\hbox{\rlap{$<$}\lower6pt\vbox{\hbox{$\sim$}}}}\ }}

\begin{document}

\title{Semi-inclusive radiative decays of $\Upsilon (1S)$}
\author{Xavier \surname{Garcia i Tormo}}
\author{Joan Soto}
\thanks{Member of CER  Astrophysics, Particle Physics and Cosmology, associated with Institut de Ci\`encies de
l'Espai-CSIC.}
\affiliation{Departament d'Estructura i Constituents de la Mat\`eria, Universitat de Barcelona\\
Diagonal 647, E-08028 Barcelona, Catalonia, Spain}
\preprint{UB-ECM-PF 05/16}

\begin{abstract}

We discuss in detail 
the photon spectrum of
radiative $\Upsilon (1S)$ decays 
taking into account a number of results that have recently appeared in
the literature. In particular, we show how to consistently combine expressions which are valid in the upper end-point region, where NRQCD factorization breaks down, with those of the central region, where NRQCD factorization holds.
An excellent description of data is achieved, but theoretical errors are large.  

\end{abstract}

\maketitle

\section{Introduction}

Semi-inclusive radiative decays of heavy quarkonium systems (see \cite{Brambilla:2004wf} for a review) to light hadrons have been a subject of investigation since the
early days of QCD \cite{Brodsky:1977du,Koller:1978qg}. It was thought for some time that a reliable extraction of $\als$
was possible from the photon spectrum normalized, for instance, to the decay into muon pairs. However, the upper
end-point region of the spectrum (namely $z\rightarrow 1$, $z$ being the fraction of the maximum energy the photon may
have) appeared to be poorly described by the theory even when Sudakov resummations were carried out
\cite{Photiadis:1985hn}. This led to some authors to claim that a non-vanishing gluon mass was necessary in order to
describe the data \cite{Consoli:1993ew}, even when relativistic corrections were taken into account \cite{Keung:1982jb}.
Later on, with the advent  of Non-Relativistic QCD (NRQCD) \cite{Bodwin:1994jh}, these decays could be analyzed in a
framework where short distance effects, at the scale of the heavy quark mass $m$ or larger, could be separated in a
systematic manner \cite{Maltoni:1998nh}. These short distance effects are calculated perturbatively in $\als (m)$ and
encoded in matching coefficients whereas long distance effects are parameterized by matrix elements of local NRQCD
operators. Even within this framework, a finite gluon mass seemed to be necessary to describe data
\cite{Consoli:1997ts}. However, about the same time it was pointed out that in the upper end-point region the NRQCD
factorization approach breaks down and shape functions, namely matrix elements of non-local operators, rather than NRQCD
matrix elements, must be introduced \cite{Rothstein:1997ac} . Early attempts to modeling color octet shape functions
produced results in complete disagreement with data \cite{Wolf:2000pm}, and hence later authors did not include them in
their phenomenological analysis \cite{Fleming:2002rv,Fleming:2002sr}.  
Notwithstanding this region has received considerable attention lately, as it was recognized that the so called
Soft-Collinear Effective Theory (SCET)\cite{Bauer:2000ew,Bauer:2000yr} may help in organizing the calculation and in
performing resummations of large (Sudakov) logs \cite{Bauer:2001rh,Fleming:2002rv,Fleming:2002sr,Fleming:2004rk}. In
fact, the early resummation of Sudakov logarithms \cite{Photiadis:1985hn} has been recently corrected
\cite{Fleming:2004rk} within this framework, and statements about the
absence of Sudakov suppression in the color singlet
channel \cite{Hautmann:2001yz} have been
clarified \cite{Fleming:2002sr}. For the $\Upsilon (1S)$ state, the bound state dynamics is amenable of a
weak coupling analysis, at least as far as the soft scale ($mv$, $v\sim \als (mv) << 1$, the typical velocity of the
heavy quark in the quarkonium rest frame) is concerned \cite{yndurain,Pineda:2001zq,sumino,hamburg,pinedahamburg,currenthamburg}.
These calculations can most conveniently be done in the framework of potential NRQCD (pNRQCD), a further effective
theory where the contributions due to the soft and ultrasoft ($\sim mv^2$) scales are factorized \cite{
Pineda:1997bj,Kniehl:1999ud,Brambilla:1999xf} (see \cite{Brambilla:2004jw} for a review). Recently a calculation of the color octet shape functions, which combines SCET and
pNRQCD, 
has become available \cite{GarciaiTormo:2004jw}.

It is the aim of this work to put together all known theoretical ingredients for these decays in order to see if a good 
description of data is achieved in the whole range of $z$, without the introduction of a
finite gluon mass
\cite{Field:2001iu}. 
The theoretical calculation of the so called direct
contributions is under good parametric control in the central region and in most of the upper end-point region. Indeed,
if one uses the original NRQCD velocity counting rules \cite{Bodwin:1994jh}, namely $\als (m)\sim v^2$ and $\als
(mv)\sim v$, together with existing calculations at weak coupling, a complete NLO expression can be put forward in the
central region. For the upper end-point region a complete LO expression, which includes color octet contributions and
takes into account both Sudakov and Coulomb resummations, is also available. The merging of the central and the
upper end-point regions will be discussed in detail. 
The fragmentation contributions, i. e. those for which the photon originates from the decay products of the heavy
quarks, 
are parametrically of the same order as the direct photon contributions in the central region \cite{Catani:1994iz} and
overweight the direct photon contributions in the lower end-point region ($z\rightarrow 0$).
They play a minor, but non-negligible, role in our analysis and are subject to large theoretical
uncertainties. 

We distribute the paper as follows. In the next section we separate the contributions to the decay width into direct and
fragmentation. Sections \ref{secdirect} and \ref{secfrag} are devoted to 
either contributions respectively. In section \ref{secphen} we carry out the phenomenological analysis and section
\ref{secconcl} is devoted to the conclusions.

\section{The photon spectrum}

The contributions to the decay width can be split into direct ($^{dir}$) and fragmentation ($^{frag}$)

\be
\frac{d\Gamma}{dz}=\frac{d\Gamma^{dir}}{dz}+\frac{d\Gamma^{frag}}{dz}
\ee
We will call direct contributions to those in which the observed photon is emitted from the heavy quarks and fragmentation contributions to those in which it is emitted from the decay products (light quarks). This splitting is correct at the order we are working but should be refined at higher orders.  $z\in [0,1]$ is defined as $z=2E_\gamma /M$ ($M$ is the mass of the heavy quarkonium state), namely the fraction of the maximum energy the photon may have in the heavy quarkonium rest frame.

\section{Direct Contributions}\label{secdirect}

The starting point is the QCD formula \cite{Rothstein:1997ac}
\begin{equation}
{d \Gamma^{dir}\over dz}=z{M\over 16\pi^2} {\rm Im} T(z)\quad \quad T(z)=-i\int d^4 x e^{-iq\cdot x}\left<
V_Q (nS)
 \vert T\{ J_{\mu} (x) J_{\nu} (0)\} \vert
V_Q (nS)
 \right> \eta^{\mu\nu}_{\perp}
\label{gdz}
\end{equation}
where $J_{\mu} (x)$ is the electromagnetic current for heavy quarks in QCD and we have restricted ourselves to $^3S_1$
states.
$q$ is the photon momentum, which in the rest frame of the heavy quarkonium is $q=\left(q_{+},q_{-},
q_{\perp}\right)=(zM,0,0)$. We have used light cone coordinates $q_\pm=q^0\pm
q^ 3$. The approximations required to calculate (\ref{gdz}) are different in
the lower end-point region ($z\rightarrow 0$), in the central region ($z\sim
0.5$) and in the upper end-point region ($z\rightarrow 1$). We will denote $\Gamma^{dir}$ by $\Gamma^{c}$ and $\Gamma^{e}$ in the central and upper end-point regions respectively (the expressions for the lower end-point region will not be necessary).

\subsection{The central 
region}

For $z$ away from the lower and upper end-points ($0$ and $1$ respectively), no further scale is introduced beyond those
inherent of the non-relativistic system.  The integration of the scale $m$ 
in the time ordered 
product of currents in (\ref{gdz}) leads to local NRQCD operators with matching coefficients which depend on $m$ and
$z$. At leading order one obtains
\begin{equation}
\label{LOrate}
\frac1{\Gamma_0} \frac{d\Gamma^c
_{\rm LO}}{dz} =  
\frac{2-z}{z} + \frac{z(1-z)}{(2-z)^2} + 2\frac{1-z}{z^2}\ln(1-z) - 2\frac{(1-z)^2}{(2-z)^3} \ln(1-z),
\end{equation}
where 
\begin{equation}
\Gamma_0 = \frac{32}{27}\alpha\alpha_s^2e_Q^2
\frac{\langle  V_Q (nS)\vert {\cal O}_1(^3S_1)\vert V_Q (nS)\rangle}{m^2},
\label{gamma0}
\end{equation}
and $e_Q$
is the charge of the heavy quark. The $\alpha_s$ correction to this rate was calculated numerically in
ref.~\cite{Kramer:1999bf}.  The expression corresponding to (\ref{gamma0}) in pNRQCD is obtained at lowest order in any
of the possible regimes by just making the substitution 
\begin{eqnarray}
\label{singletWF}
\langle  V_Q (nS) \vert {\cal O}_1(^3S_1) \vert  V_Q (nS) \rangle &=&
2 N_c |\psi_{n0}({\bf 0})|^2,
\end{eqnarray}
where 
$\psi_{n0}({\bf 0})$ is the wave function at the origin. The final result coincides with the one of the early QCD
calculations \cite{Brodsky:1977du,Koller:1978qg}. We will take the Coulomb form 
$\psi_{10}({\bf 0})=\gamma^3/\pi$ for the LO analysis of $\Upsilon (1S)$ ($\gamma$ is defined in (\ref{defg})). 

The NLO contribution in the original NRQCD counting
\cite{Bodwin:1994jh} is $v^2$ suppressed with respect to (\ref{LOrate}). It reads

\begin{equation}
\label{RelCo}
\frac{d\Gamma^c_{\rm NLO}}{dz}=C_{\mathbf{1}}'\left(\phantom{}^3S_1\right)\frac{\langle  V_Q (nS)\vert {\cal
P}_1(^3S_1)\vert V_Q (nS)\rangle}{m^4}
\end{equation}

In the original NRQCD counting or in the weak coupling regime of pNRQCD the new matrix element above can be written in
terms of the original one \cite{Gremm:1997dq}\footnote{In the strong coupling regime of pNRQCD an additional contribution appears \cite{Brambilla:2002nu}}

\begin{equation}
\frac{\langle  V_Q (nS)\vert {\cal P}_1(^3S_1)\vert V_Q (nS)\rangle}{m^4}=\left(\frac{M-2m}{m}\right)\frac{\langle  V_Q
(nS)\vert {\cal O}_1(^3S_1)\vert V_Q (nS)\rangle}{m^2}\left(1+\mathcal{O}\left(v^2\right)\right)
\end{equation}

The matching coefficient can be extracted from an early calculation \cite{Keung:1982jb} (see also \cite{Yusuf:1996av}).
It reads

\begin{equation}
C_{\mathbf{1}}'\left(\phantom{}^3S_1\right)=-\frac{16}{27}\alpha\alpha_s^2e_Q^2\left(F_B(z)+\frac{1}{2} F_W(z)\right)
\end{equation}
where ($\xi=1-z$)

\[
F_B(z)=\frac{2-16\xi+10\xi^2-48\xi^3 -10\xi^4+64\xi^5-2\xi^6 +(1-3\xi+14\xi^2-106\xi^3+17\xi^4 -51\xi^5)\ln
\xi}{2\,(1-\xi)^3 
(1+\xi)^4}
\]
\[
F_W(z)=\frac{-26+14\xi-210\xi^2+134\xi^3+274\xi^4-150\xi^5-38\xi^6+2\xi^7}{3(1-\xi)^3 (1+\xi)^5}-
\]
\begin{equation}
-\frac{(27+50\xi+257\xi^2-292\xi^3+205\xi^4-78\xi^5-41\xi^6)\ln \xi}{3(1-\xi)^3 (1+\xi)^5}
\end{equation}

The contributions of color octet operators start at order $v^4$. Furthermore, 
away of the upper end-point region, the lowest order color octet contribution identically vanishes
\cite{Maltoni:1998nh}. Hence there is no $1/\als$ enhancement in the central region and we can safely neglect these
contributions here.

If we use the counting $\als (\mu_h)\sim v^2$, $\als\left(\mu_s\right)\sim v$ ($\mu_h\sim m$ and $\mu_s\sim mv$ are the hard and the soft scales 
respectively) for the $\Upsilon (1S)$, the complete result up to NLO (including $v^2$ suppressed contributions) can be written as
\be
\frac{d\Gamma^c}{dz}=\frac{d\Gamma^{c}_{LO}}{dz}+\frac{d\Gamma^{c}_{NLO}}{dz}+\frac{d\Gamma^{c}_{LO,\als}}{dz}
\label{central}
\ee
The first term consist of the expression (\ref{LOrate}) with the Coulomb wave
function at the origin (\ref{singletWF}) including corrections up to
$\mathcal{O}\left[\left(\als\left(\mu_s\right)\right)^2\right]$
\cite{Melnikov:1998ug,Penin:1998kx}, the second term is given in
(\ref{RelCo}), and the third term consists of the radiative $\mathcal{O}\left(\als (\mu_h)\right)$ corrections 
to (\ref{LOrate}) which have been calculated numerically in \cite{Kramer:1999bf}. Let us mention
at this point that the $\mathcal{O}\left[\left(\als\left(\mu_s\right)\right)^2\right]$ 
corrections to the wave function at the origin turn out to be as large as the leading order 
term. This will be important for our final results.  Note that the standard
NRQCD counting we use does not coincide with the usual counting of pNRQCD in
weak coupling calculations, where $\als (\mu_h) \sim \als (\mu_s) \sim \als
(mv^2)$.  The latter is necessary in order to get factorization scale
independent results beyond NNLO for the spectrum and beyond NLO for creation
and annihilation currents. However, for the $\Upsilon (1S)$ system (and the
remaining heavy quarkonium states) the ultrasoft scale $mv^2$ is rather low,
which suggests that perturbation theory should better be avoided at this scale
\cite{Pineda:2001zq}. This leads us to standard NRQCD counting. The factorization scale dependences that this counting induces can in principle be avoided using renormalization group techniques \cite{Luke:1999kz,Pineda:2001ra,Pineda:2001et,Pineda:2002bv,Hoang:2002yy}. In practice, however, only partial NNLL results exists for the creation and annihilation currents \cite{Hoang:2003ns,Penin:2004ay} (see \cite{Pineda:2003be} for the complete NLL results), which would fix the scale dependence of the  wave function at the origin at ${\cal O} (\als^2 (mv))$. We will not use them and will just set the factorization scale to $m$. 
\subsection{The 
lower end-point region}

For $z\rightarrow 0$, the emitted low energy photon can only produce transitions within the non-relativistic bound state
without destroying it. Hence the direct low energy photon emission takes place in two steps: (i) the photon is emitted
(dominantly by dipole electric and magnetic transitions) and (ii) the remaining (off-shell) bound state is annihilated
into light hadrons. It has a suppression  $\sim z^ 3$ with respect to $\Gamma_0$ (see
\cite{Manohar:2003xv,Voloshin:2003hh} for a recent analysis of this region in QED). 
 Hence, at some point the direct photon emission is overtaken by the fragmentation contributions 
\cite{Catani:1994iz,Maltoni:1998nh}. In practice this happens about $z\sim 0.4$,
namely much before than the $z^3$ behavior of the low energy direct photon emission can be observed, and hence we shall
neglect the latter in the following.

\subsection{The 
upper end-point region}

In this region the standard NRQCD factorization is not applicable \cite{Rothstein:1997ac}. This is due to the fact that
small scales induced by the kinematics enter the problem and have an interplay with the bound state dynamics. In order
to study this region, one has to take into account collinear degrees of freedom in addition to those of NRQCD. This can
be done using SCET as it has been described in \cite{Bauer:2001rh,Fleming:2002sr}. 
In this region, the color octet contributions are only suppressed by $v^2$ or by $1-z$. Since their matching
coefficients are enhanced by $1/\als (\mu_h)$, they become as important as the color singlet contributions if we count $\als
(\mu_h)\sim v^2\sim 1-z$. 
We will write
\be
\frac{d\Gamma^e}{dz}=\frac{d\Gamma^{e}_{CS}}{dz}+\frac{d\Gamma^{e}_{CO}}{dz}
\label{endp}
\ee
where $CS$ and $CO$ stand for color singlet and color octet contributions respectively.

\subsubsection{
Color singlet contributions}

For the color singlet contribution we shall use the expression with the Sudakov resummed coefficient in ref.
\cite{Fleming:2004rk}
\[
\frac{1}{\Gamma_0}\frac{d\Gamma^{e}_{CS}}{dz} = \Theta(M-2mz) \frac{8z}9 
\sum_{n \rm{\ odd}} \left\{\frac{1}{f_{5/2}^{(n)}}
\left[ \gamma_+^{(n)} r(\mu_c)^{2 \lambda^{(n)}_+ / \beta_0}  - 
\gamma_-^{(n)} r(\mu_c)^{2 \lambda^{(n)}_- / \beta_0} \right]^2+\right.
\]
\begin{equation}
\label{singres}
\left.+
\frac{3 f_{3/2}^{(n)}}{8[f_{5/2}^{(n)}]^2}\frac{{\gamma^{(n)}_{gq}}^2}{\Delta^2}
\left[ r(\mu_c)^{2 \lambda^{(n)}_+ / \beta_0}  -  
r(\mu_c)^{2 \lambda^{(n)}_- / \beta_0} \right]^2\right\}
\end{equation}
where the definitions for the different functions appearing in (\ref{singres}) are collected in the Appendix \ref{appdef}.

\subsubsection{
Color octet contributions}\label{subsubsecCO}

For the color octet contributions we use
\begin{equation}
\frac{d\Gamma_{CO}^{e}}{dz}=\alpha_s\left(\mu_u\right)\alpha_s\left(\mu_h\right)\left(\frac{16M\alpha}{81m^4%
}\right)\int_z^{\frac{M}{2m}}\!\!\! C(x-z) S_{S+P}(x)dx
\end{equation}
$\mu_u$ is the ultrasoft (US) scale (see section \ref{secphen} for the expression we use). $C(x-z)$ contains 
the Sudakov resummations of ref. \cite{Bauer:2001rh}\footnote{The matching coefficients provided in 
this reference become imaginary for extremely small values of $z-1$, a region where our results do not hold anyway. We
have just cut-off this region in the convolutions.},
\begin{equation}\label{ocres} C(x-z)=-\frac{d}{dz} \left\{
\theta(x-z) \; \frac{\exp [ \ell g_1[\alpha_s \beta_0 \ell/(4\pi)] + g_2[\alpha_s \beta_0
\ell/(4\pi)]]}{\Gamma[1-g_1[\alpha_s \beta_0
\ell/(4\pi)] - \alpha_s \beta_0 \ell/(4\pi) g_1^\prime[\alpha_s \beta_0
\ell/(4\pi)]]}\right\}
\end{equation}
 The (tree level) matching coefficients (up to a global factor) and the various shape functions are encoded in $S_{S+P}(x)$,
\begin{equation}
S_{S+P}(z):=z\left(-\left(\frac{4\alpha_s\left(\mu_u\right)}{3\pi
N_c}\left(\frac{c_F}{2m}\right)^2\right)^{-1}\!\!\!\!\!S_S(M(1-z))-\left(\frac{\alpha_s\left(\mu_u\right)}{6\pi
N_c}\right)^{-1}\left(3S_{P1}(M(1-z))+S_{P2}(M(1-z))\right)\right)
\label{sp}
\end{equation}
The definitions of all the functions appearing in (\ref{ocres}) and (\ref{sp})
are collected in the Appendix \ref{appdef}. The shape functions $S_S$, $S_{P1}$ and $S_{P2}$ may become
$S_S^{MS}$, $S_{P1}^{MS}$ and $S_{P2}^{MS}$ or  $S_S^{sub}$, $S_{P1}^{sub}$ and $S_{P2}^{sub}$ 
depending on the subtraction scheme employed. The procedure used to renormalize
the shape functions is explained in the Appendix \ref{appregren}. 
In figure
\ref{epsubst} we plot the end-point contribution (\ref{endp}) with the shape
functions renormalized in an $MS$ scheme (blue dashed line) and in the $sub$ scheme, which makes
 additional subtractions (red solid line), together with the experimental
data \cite{Nemati:1996xy} (we have convoluted the theoretical curves with the
experimental efficiency, the overall normalization of each curve is taken
as a free parameter. For the details of the scale setting see section \ref{secphen}). We see that both schemes are equally good for the description of the shape of the experimental data in the end-point region.

Note (from Appendix A) that we use the octet shape functions calculated in ref. \cite{GarciaiTormo:2004jw}, which are crucial in order to have a good
description of data in the upper end-point region. We would like to comment on the validity of those formulas. This is
limited by the perturbative treatment of the US gluons. The typical momentum
of these gluons in light cone
coordinates turns out to be:
\begin{equation}
(k_+, k_{\perp}, k_-)=\left(M(1-z),\sqrt{2M(1-z)\left(\frac{M(1-z)}{2}-E_1\right)},M(1-z)-2E_1\right)
\end{equation}
Note that the typical $k_{\perp}$ is not fixed by the bound state dynamics only but by a combination of the latter and
the end-point kinematics. Hence, the calculation is reliable provided that $k_{\perp} \gtrsim 1 GeV.$, which means $z <
0.92$. Note also that the typical three momentum of the heavy quarks in the shape function is given by
\begin{equation}
p\sim\sqrt{m\left(\frac{M}{2}(1-z)-E_1\right)}
\end{equation}
This means that the multipole expansion always holds in the end-point region, 
regardless that $M(1-z)$ is bigger or smaller than $m\als^2$. This important point was not sufficiently emphasized in
\cite{GarciaiTormo:2004jw} and it is the ultimate reason why such a good description of data was obtained there for
$z\in [ 0.7,1 ]$. 
Recall that at $z<0.7$ or so we are at the border of the end-point region and the standard NRQCD factorization formulas
should hold. This means that the contributions of the octet shape functions should merge suitable contributions of the
NRQCD factorization formulas, as we discuss below.

\begin{figure}
\centering
\includegraphics[width=15cm]{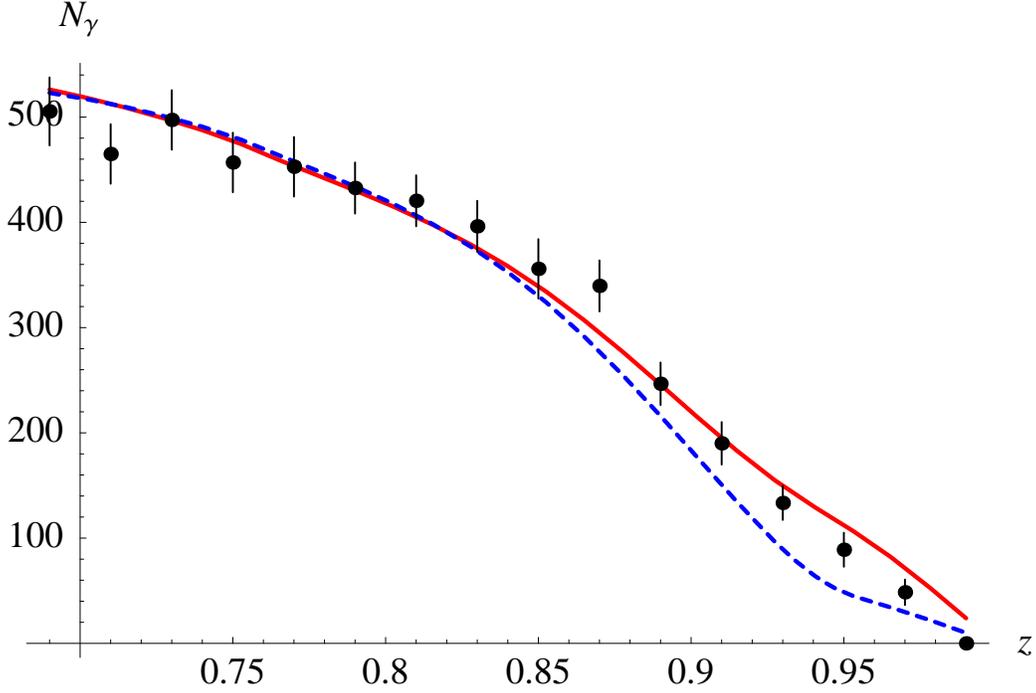}
\caption{End-point contribution of the spectrum, 
$d\Gamma^e/dz$, with
  the shape functions renormalized in an $MS$ scheme (blue dashed line) and
  in the $sub$ scheme 
(red solid line). The points are the CLEO data
  \cite{Nemati:1996xy}}\label{epsubst}
\end{figure}

\subsection{Merging the central and 
upper end-point regions}\label{subsecmatch}

As we have seen, different approximations are necessary in the central and upper end-point regions. It is then not obvious how the results for the central
and for the upper end-point regions must be combined in order to get a reliable description of the whole spectrum.
When the results of the central region are used in the upper end-point region, one misses certain Sudakov and Coulomb resummations which are necessary because the softer scales $M\sqrt{1-z}$ and $M(1-z)$ become relevant. Conversely, when results for the end-point region are used in the central region, one misses non-trivial functions of $z$, which are approximated by their end-point ($z\sim 1$) behavior.

One way to proceed is the following. If we assume that the expressions for the end-point contain the ones of the central region up to a certain order in $(1-z)$, 
we could just subtract from the expressions in the central region the behavior
when $z\rightarrow 1$ at the desired order and add the expressions in the
end-point region. Indeed, when $z\rightarrow 1$ this procedure would improve
on the central region expressions up to a given order in $(1-z)$, and when $z$
belongs to the central region, they would reduce to the central region
expressions up to higher orders in $\als$. This method was used in ref. \cite{Fleming:2002sr} and in ref. \cite{GarciaiTormo:2004kb}. In ref. \cite{Fleming:2002sr} only color singlet contributions were considered and the end-point expressions trivially contained the central region expressions in the limit $z\rightarrow 1$. In ref. \cite{GarciaiTormo:2004kb} color octet contributions were
included, which contain terms proportional to $(1-z)$. Hence, the following formula was used
\begin{equation}
\frac{1}{\Gamma_0}\frac{d\Gamma^{dir}}{dz}=\frac{1}{\Gamma_0}\frac{d\Gamma_{LO}^{c}}{dz}+\left(\frac{1}{\Gamma_0%
}\frac{d\Gamma_{CS}^{e}}{dz}-z\right)+\left(\frac{1}{\Gamma_0}\frac{d\Gamma_{CO}^{e}}{dz}-z\left(4+2\log \left(1-z\right)
\right) (1-z)\right)
\label{mergingLO}
\end{equation}
Even though a remarkable description of data was achieved with this formula (upon using a suitable subtraction scheme described below), this method suffers from the following shortcoming. The hypothesis that the expressions for the end-point contain the ones for the central region up to a given order in $(1-z)$ is in general not fulfilled. As we will see below, typically, they only contain part of the expressions for the central region. This is due to the fact that some $\als (\mu_h)$ in the central region may soften as $\als (M(1-z))$, others as $\als (M\sqrt{1-z})$ and others may stay at $\als (\mu_h)$ when approaching the end-point region. In a LO approximation at the end-point region, only the terms with the $\als$ at low scales would be kept and the rest neglected, producing the above mentioned mismatch. We shall not pursue this procedure any further.

Let us look for an alternative.  
Recall first that
the expressions we have obtained for the upper end-point region are non-trivial functions of $M(1-z)$, $M\sqrt{1-z}$, $m\als (mv)$ and $m\als^2 (mv)$, which involve $\als$ at all these scales. They take into account both Sudakov and Coulomb resummations. 
When $z$ approaches the central region, we can expand them in $\als (M\sqrt{1-z})$, $\als (M(1-z))$ and the ratio $m\als (mv)/M\sqrt{1-z}$. They should reduce to the form of the expressions for the central region, since we are just undoing the Sudakov and (part of) the Coulomb resummations. Indeed, we obtain
\bea\label{expand} 
\frac{d\Gamma^{e}_{CS}}{dz} &\longrightarrow  \displaystyle{\left.\frac{d\Gamma^{e}_{CS}}{dz}\right\vert_c}= & \Gamma_0z\left(1+\frac{\als}{6\pi}\left(C_A\left(2\pi^2-17\right)+2n_f\right)\log (1-z) + \mathcal{O}(\als^2) \right)\label{expands}\\
\frac{d\Gamma^{e}_{CO}}{dz} &\longrightarrow  \displaystyle{\left.\frac{d\Gamma^{e}_{CO}}{dz}\right\vert_c} = & -z\als^2\left(\frac{16M\alpha}{81m^4}\right)2\left|\psi_{10}\left(\bf{0}\right)\right|^2
\bigg( m\als\sqrt{1-z} A
+\nonumber\\
 & & \left.+
M(1-z)\left(-1+\log\left(\frac{\mu_c^2}{M^2(1-z)^2}\right)\right)+\right.\nonumber\\
 & & \left.+
M\frac{\als}{2\pi}\left(-2C_A\left(\frac{1}{2}(1-z)\log^2(1-z)\left[\log\left(\frac{\mu_c^2}{M^2(1-z)^2}\right)-1\right]+\int_z^1\!\!\!dx\frac{\log(x-z)}{x-z}f(x,z)\right)-\right.\right.\nonumber\\
 & & \left.\left.-\left(\frac{23}{6}C_A-\frac{n_f}{3}\right)\left((1-z)\log(1-z)\left[\log\left(\frac{\mu_c^2}{M^2(1-z)^2}\right)-1\right]+\int_z^1\!\!\!dx\frac{1}{x-z}f(x,z)\right)\right)-\right.\nonumber\\
 & & \left.-
\frac{\gamma^2}{m}2\left(\log\left(\frac{\mu_c^2}{M^2(1-z)^2}\right)+1\right)+ \mathcal{O}\left( m\als^2, \als \frac{\gamma^2}{m}, \frac{\gamma^4}{m^3}\right) \right)\label{expando}
\eea
where
\begin{equation}
f(x,z)=\left(1-x\right)\log\left(\frac{\mu_c^2}{M^2(1-x)^2}\right)-(1-z)\log \left(\frac{\mu_c^2}{M^2(1-z)^2}\right)+x-z
\end{equation}
$A=-N_c-136C_f(2-\lambda)/9$ (in an $MS$ scheme; it becomes $A=-64C_f(2-\lambda)/9$ in the 
$sub$
scheme described in Appendix B). The details of this derivation are given in the 
Appendix \ref{appcenreg}. The color singlet contribution reproduces the full LO expression 
for the central region in the limit $z\rightarrow 1$. 
The color octet shape functions $S_{P1}$  and $S_{P2}$ give contributions to the
relativistic corrections (\ref{RelCo}), and $S_{P2}$ to terms
proportional to $(1-z)$ in the limit $z\rightarrow 1$ of (\ref{LOrate}) as well. We
have checked that, in the $z\rightarrow 1$ limit, both the $(1-z)\ln (1-z)$ of
(\ref{LOrate}) and the $\ln (1-z)$ of the relativistic correction
(\ref{RelCo}) are correctly reproduced if $\mu_c \sim M\sqrt{1-z}$, as it
should. 
All the color octet shape functions 
contribute to the $\mathcal{O}(\als (\mu_h))$
correction in the
first line of (\ref{expando}).
There are additional $\mathcal{O}(\als (\mu_h))$ contributions coming from the expansion of
the (Sudakov) resummed matching coefficients of the color singlet contribution
and of the $S_{P2}$ color octet shape function. The $\als\log(1-z)$ in
(\ref{expands}) 
reproduces the logarithm in ${d\Gamma^{c}}_{LO,\als}/{dz}$.

We propose the following formula
\be
\frac{1}{\Gamma_0}\frac{d\Gamma^{dir}}{dz}=\frac{1}{\Gamma_0}\frac{d\Gamma^{c}}{dz}+\left(\frac{1}{\Gamma_0
}\frac{d\Gamma_{CS}^{e}}{dz}-\left.{\frac{1}{\Gamma_0
}\frac{d\Gamma_{CS}^{e}}{dz}}\right\vert_c\right)+\left(\frac{1}{\Gamma_0}\frac{d\Gamma_{CO}^{e}}{dz}-\left.{\frac{1}{\Gamma_0
}\frac{d\Gamma_{CO}^{e}}{dz}}\right\vert_c\right)
\label{mergingNLO}
\ee
This formula reduces to the NRQCD expression in the central
region. When we approach the upper end-point region the second terms in each
of the parentheses are expected to cancel corresponding terms in the $z\to1$
limit of the expression for the central region up to higher order terms (in the end-point region counting). Thus, we are left with the resummed
expressions for the end-point (up to higher order terms). 

There are of course other possibilities for the merging. For instance, one may
choose a $z_1$ below which one trusts the calculation for the central region
and a $z_2$ above which one trusts the end-point region calculation, and use
some sort of interpolation between $z_1$ and $z_2$ (see for instance
\cite{Lin:2004eu}). This would have the advantage of keeping the right
approximation below $z_1$ and beyond $z_2$ unpolluted, at the expense of
introducing further theoretical ambiguities due to the choice of $z_1$ and
$z_2$, and, more important, due to the choice of the interpolation between
$z_1$ and $z_2$. We believe that our formula (\ref{mergingNLO}) is superior
because it does not introduce the above mentioned theoretical ambiguities. The
price to be paid is that the expressions from the central region have an influence in the end-point region and vice-versa. This influence can always be chosen to be parametrically subleading but large numerical factors may make it noticeable in some cases, as we shall see below.

\subsubsection{Merging at LO}\label{subsubsecLO}

\begin{figure}
\centering
\includegraphics[width=15cm]{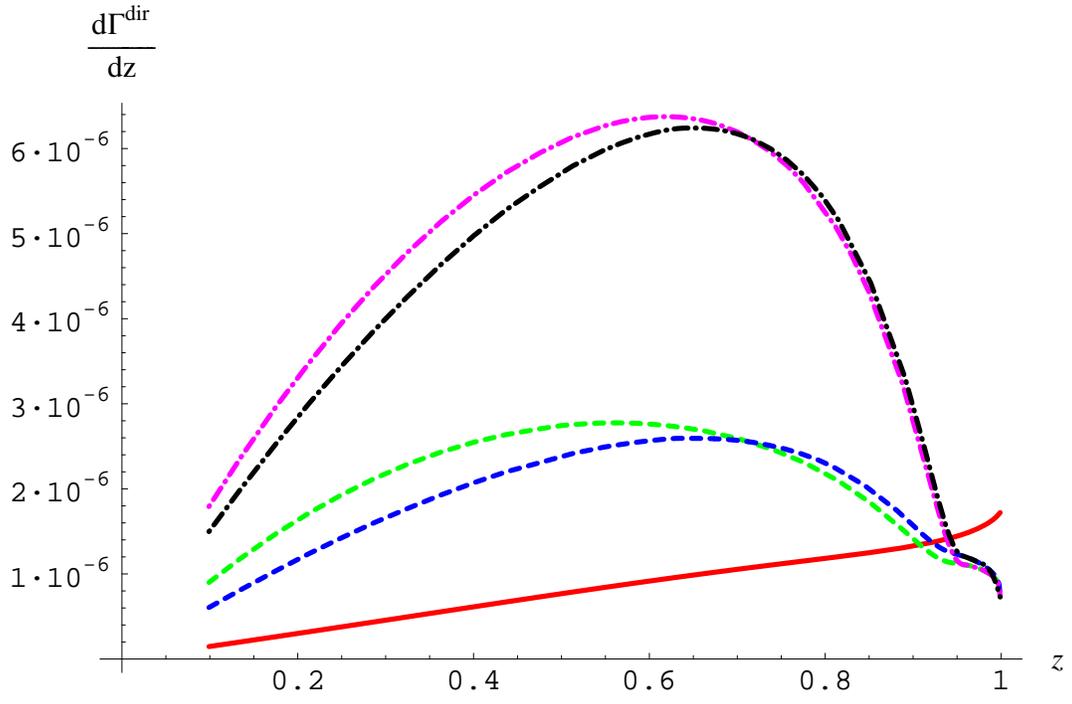}
\caption{Merging at LO. The solid red line is the NRQCD expression
  (\ref{LOrate}). The dot-dashed curves are obtained using an $MS$ scheme: the
  pink (light) curve is the end-point contribution (\ref{endp}) and the
  black (dark) curve is the LO merging of section \ref{subsubsecLO}. The dashed
  curves are obtained using the $sub$ scheme 
(explained
  in the Appendix \ref{appregren}): the green (light) curve is the end-point contribution (\ref{endp}) and the blue (dark) curve is the LO merging of
  section \ref{subsubsecLO}.}\label{compLO}
\end{figure}

\begin{figure}
\centering
\includegraphics[width=15cm]{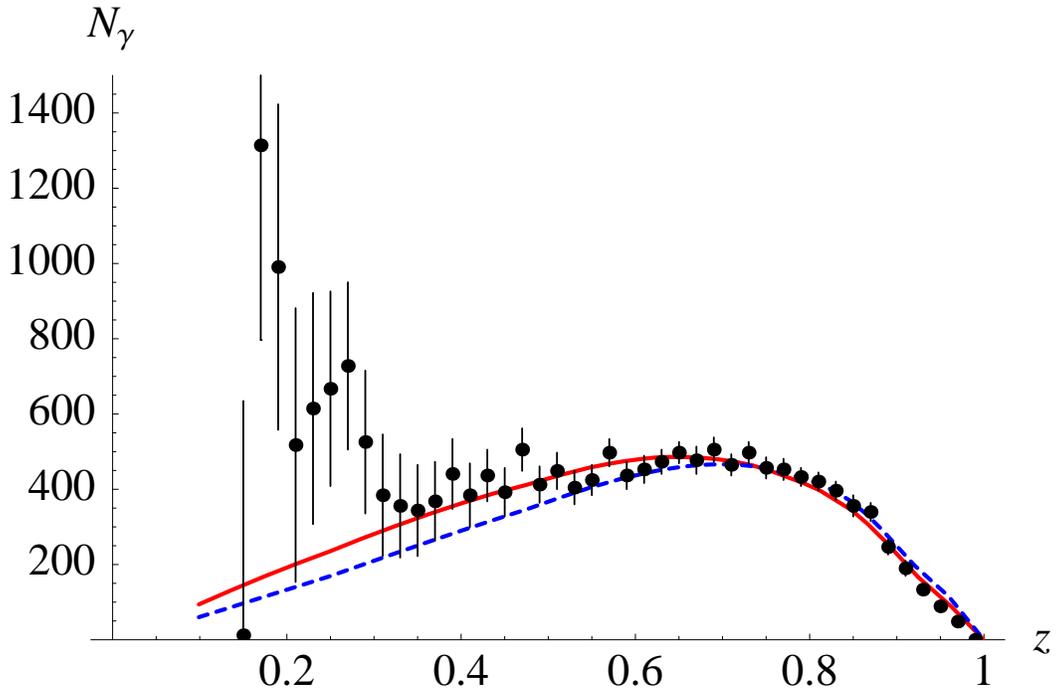}
\caption{Direct contribution to the spectrum. The solid red line corresponds
  to the LO merging of section \ref{subsubsecLO} and the blue dashed line
  corresponds to equation (\ref{mergingLO}). The points are the CLEO data
  \cite{Nemati:1996xy}.}\label{dibLO}
\end{figure}

If we wish to use only the LO expressions for the central region, we should take (\ref{expando}) at LO, namely
\be
\left. { \frac{1}{\Gamma_0
}\frac{d\Gamma_{CS}^{e}}{dz}}\right\vert_c = z \quad ,\quad \left. {\frac{1}{\Gamma_0
}\frac{d\Gamma_{CO}^{e}}{dz}}\right\vert_c= z\left(2-4\log \left(\frac{\mu_c}{M(1-z)}\right)
\right) (1-z) 
\ee
and substitute them in (\ref{mergingNLO}).
Unexpectedly, the results obtained with this formula in the central region deviate 
considerably from those obtained with formula (\ref{LOrate}) (see Fig. \ref{compLO}). This can be
traced back to the fact that the $\als\sqrt{1-z} $ corrections in (\ref{expando}) are enhanced 
by large numerical factors, which indicates that the merging should better be done including 
$\als (\mu_h)$ corrections in the central region, as we discuss in the next section. 
Alternatively,
we may change our subtraction scheme in order to (partially) get rid of
these contributions, as discussed in the Appendix \ref{appregren}. With the new subtraction 
scheme ($sub$) the situation improves, although it does not become fully satisfactory (see Fig. \ref{compLO}).
This is due to the fact that some $\als\sqrt{1-z} $ terms remain, which do not seem to be 
associated to the freedom of choosing a particular subtraction scheme. In spite of this the description of data turns out to be extremely
good. In
figure \ref{dibLO} we plot,
using the $sub$ scheme, 
the merging at LO (solid red line) and also, for comparison,
equation (\ref{mergingLO}) (blue dashed line).
We have convoluted
the theoretical curves with the experimental efficiency and the overall
normalization is taken as a free parameter.

\subsubsection{Merging at NLO}\label{subsubsecNLO}

\begin{figure}
\centering
\includegraphics[width=15cm]{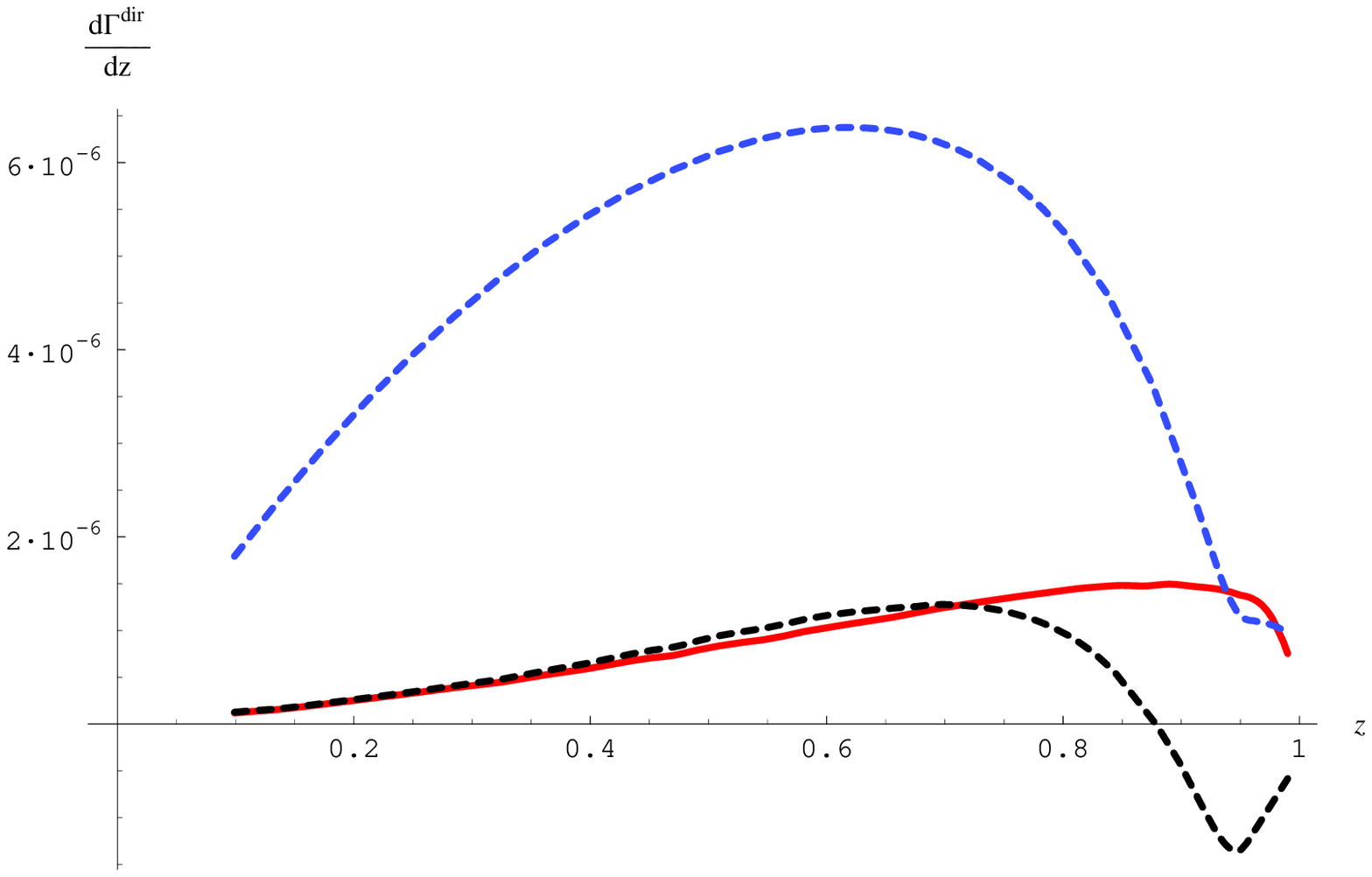}
\caption{Merging at NLO (using an $MS$ scheme and the wave function at the
  origin at LO). The solid red line is the NRQCD result
  (\ref{central}), the blue (light) dashed curve is the end-point contribution
  (\ref{endp}) and the black (dark) dashed curve is the NLO merging of section
  \ref{subsubsecNLO}.}\label{compNLOFO}
\end{figure}

\begin{figure}
\centering
\includegraphics[width=15cm]{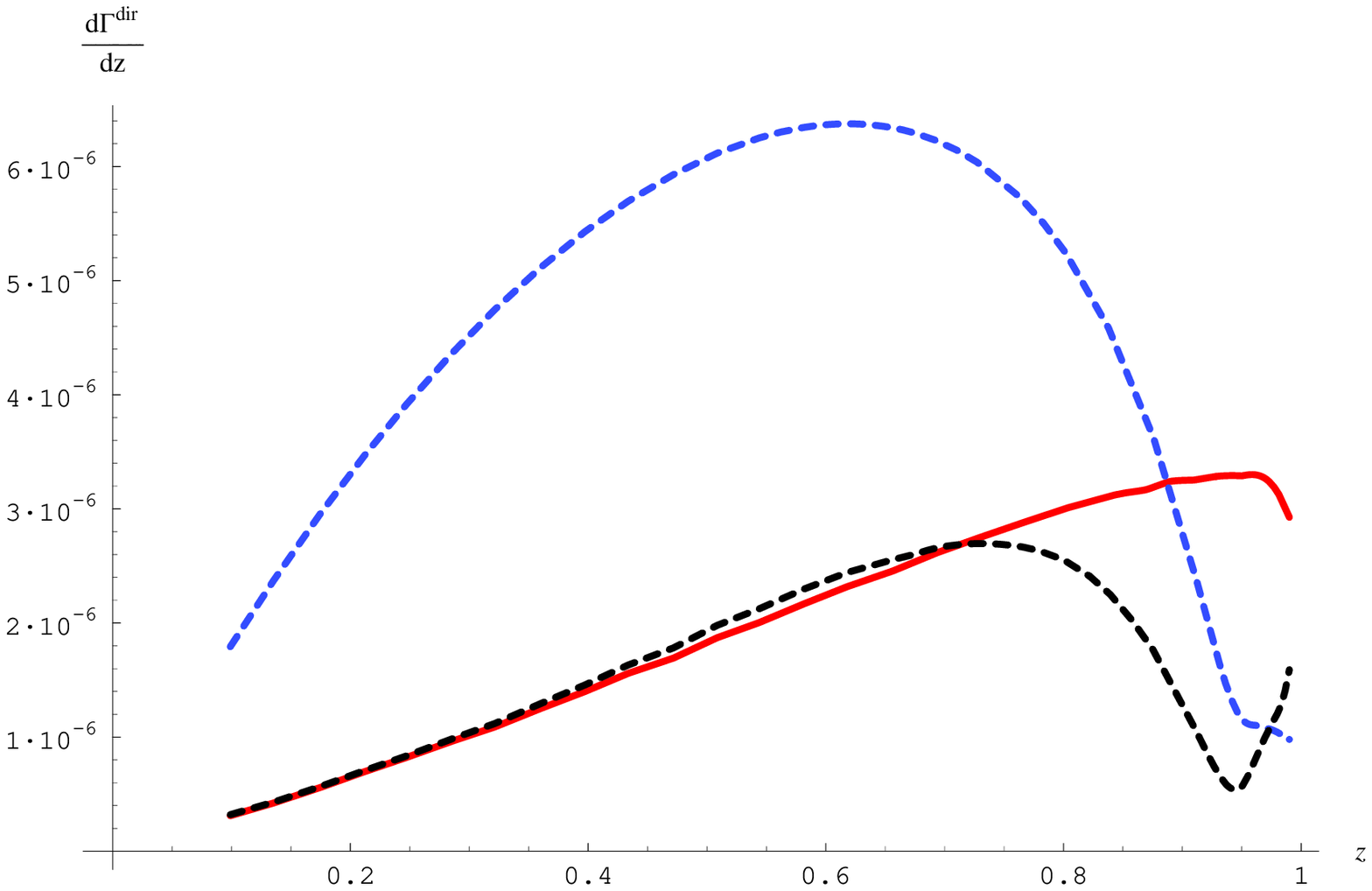}
\caption{Merging at NLO (using an $MS$ scheme and the wave function at the
  origin with the $\als^2 (\mu_s)$ corrections included). 
The solid red line is the NRQCD result
  (\ref{central}), the blue (light) dashed curve is the end-point contribution
  (\ref{endp}) and the black (dark) dashed curve is the NLO merging of section
  \ref{subsubsecNLO}.}\label{compNLOFO2}
\end{figure}

If we wish to use the NLO expressions for the central region (\ref{central}),
we should take all the terms displayed in (\ref{expands})- (\ref{expando}) and substitute them 
in (\ref{mergingNLO}).
Unlike in the LO case, for values of $z$ in the central region the curve obtained from 
(\ref{mergingNLO}) now approaches smoothly the expressions for the central region 
(\ref{central}) as it should. This is so no matter if we include the $\als^2 (\mu_s)$ corrections to 
the wave function at the origin in $d\Gamma^c_{LO}/dz$, as we in principle should, or not
 (see Figs. \ref{compNLOFO} and \ref{compNLOFO2}). However, since the above corrections are very large, the behavior of the 
curve for $z\rightarrow 1$, strongly depends 
on whether we include them or not (see again Figs. \ref{compNLOFO} and \ref{compNLOFO2}). We believe that the two possibilities 
are legitimate. If one interpretes the large $\als^2 (\mu_s)$ corrections as a sign that the asymptotic
series starts exploding, one should better stay at LO (or including $\als (\mu_s)$ corrections). However, if one believes that the large 
$\als^2 (\mu_s)$ corrections are an accident and that the $\als^3 (\mu_s)$ ones (see \cite{Beneke:2005hg,Penin:2005eu} for partial results) will again be 
small, one should use these $\als^2 (\mu_s)$ corrections. We consider below the two cases.
 
If we stay at LO (or including $\als (\mu_s)$ corrections) for the wave function at the origin, the curve we 
obtain for $z\rightarrow 1$ differs
considerably from the expressions for the end-point region (\ref{endp}) (see Fig. \ref{compNLOFO}). This can be traced 
back to 
the $\als\sqrt{1-z}$ term
in (\ref{expando}) again. This term is parametrically suppressed in the
end-point region, but, since it is multiplied by a large numerical factor, its
contribution turns out to be overwhelming.
This term might (largely) cancel out
against higher order contributions in the end-point region, in particular
against certain parts of the NLO expressions for the color singlet
contributions, which are unknown at the moment. 

\begin{figure}
\centering
\includegraphics[width=15cm]{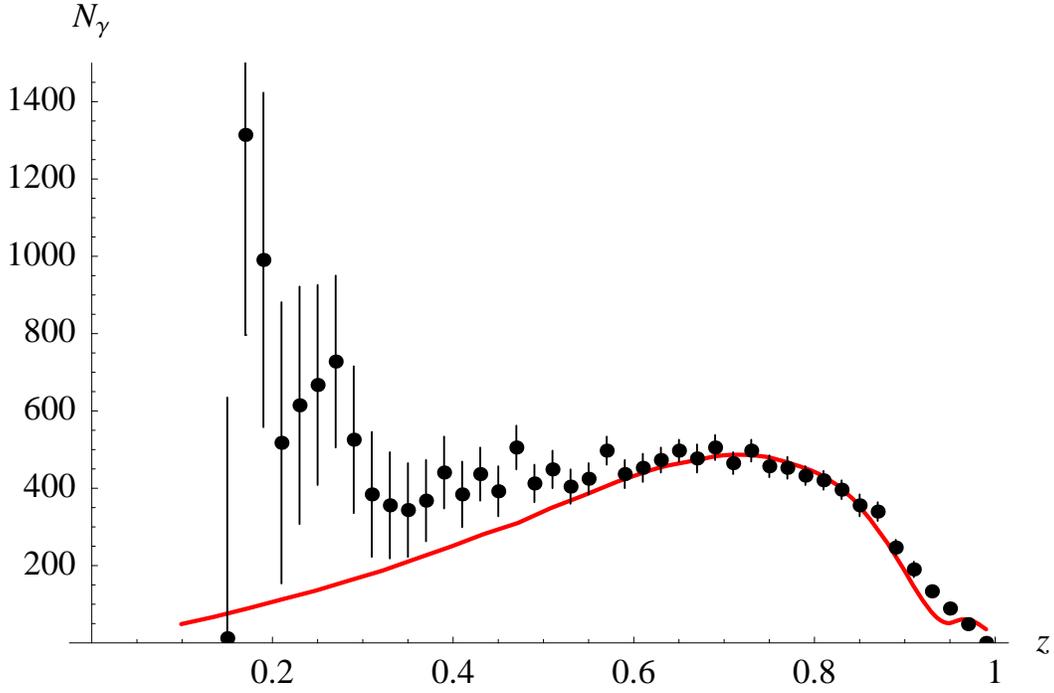}
\caption{Direct contribution to the spectrum using the NLO merging of section
  \ref{subsubsecNLO} (in an $MS$ scheme and the wave function at the
  origin with the $\als^2 (\mu_s)$ corrections included). The points are the CLEO data
  \cite{Nemati:1996xy}.}\label{dibNLO}
\end{figure}

If we use the wave function at the origin with the $\als^2 (\mu_s)$ corrections included, the curves we obtain for $z\rightarrow 1$ become much closer to the expressions for the end-point 
region (\ref{endp}) (see Fig \ref{compNLOFO2}). Hence, a good description of data is obtained 
with no need of additional subtractions, as shown in figure \ref{dibNLO} (as
usual experimental efficiency has been taken into account and the overall
normalization is a free parameter).

\section{Fragmentation contributions}\label{secfrag}
The fragmentation contributions can be written as
\begin{equation}
\frac{d\Gamma^{frag}}{dz}=\sum_{a = q,\bar q, g} \int_z^1\frac{dx}{x}C_a(x)D_{a\gamma}\left(\frac{z}{x},M\right),
\end{equation}
where $C_a$ represents the partonic kernels and $D_{a\gamma}$ represents the fragmentation functions. The partonic
kernels can again be expanded in powers of $v$ \cite{Maltoni:1998nh}
\begin{equation}
C_a=\sum_{\mathcal{Q}}C_a[\mathcal{Q}]
\end{equation}
The leading order term in $v$ is the color singlet rate to produce three gluons
\begin{equation}
\label{fragsing} C_g\left[{\cal O}_1(^3S_1)\right]=\frac{40}{81}\alpha_s^3
\left(\frac{2-z}{z} + \frac{z(1-z)}{(2-z)^2} + 2\frac{1-z}{z^2}\ln(1-z) - 2\frac{(1-z)^2}{(2-z)^3}
\ln(1-z)\right)\frac{\langle  V_Q (nS)\vert {\cal O}_1(^3S_1)\vert V_Q (nS)\rangle}{m^2}
\end{equation}
The color octet contributions start at order $v^4$ but have a $\frac{1}{\alpha_s}$ enhancement with respect to
(\ref{fragsing})
\[
C_g\left[{\cal O}_8(^1S_0)\right]=\frac{5\pi\alpha_s^2}{3}
\delta(1-z)\frac{\langle V_Q (nS)\vert{\cal O}_8(^1S_0)\vert V_Q (nS)\rangle}{m^2}
\]
\[
C_g\left[{\cal O}_8(^3P_J)\right]=\frac{35\pi\alpha_s^2}{3}
\delta(1-z)\frac{\langle V_Q (nS)\vert{\cal O}_8(^3P_0)\vert V_Q (nS)\rangle}{m^4}
\]
\begin{equation}
C_q\left[{\cal O}_8(^3S_1)\right]=\frac{\pi\alpha_s^2}{3}
\delta(1-z)\frac{\langle V_Q (nS)\vert{\cal O}_8(^3S_1)\vert V_Q (nS)\rangle}{m^2}\label{octetf}
\end{equation}

Then the color singlet fragmentation contribution is of order $\alpha_s^3D_{g\to\gamma}$ and the color
octet fragmentation are of order $v^4\alpha_s^2D_{g\to\gamma}$ ($\phantom{}^1S_0$ and $\phantom{}^3P_J$ contributions)
or $v^4\alpha_s^2D_{q\to\gamma}$ ($\phantom{}^3S_1$ contribution). We can use, as before, the counting
$v^2\sim\alpha_s$ to compare the relative importance of the different contributions together with the existing models for the
fragmentation functions \cite{Aurenche:1992yc}. The latter tell us that $D_{q\to\gamma}$ is much larger than $D_{g\to\gamma}$. 
This causes the $\mathcal{O}(v^4\alpha_s^2D_{q\to\gamma})$ $\phantom{}^3S_1$ octet contribution to dominate in front of the singlet 
$\mathcal{O}(\alpha_s^3D_{g\to\gamma})$ and the octet $\mathcal{O}(v^4\alpha_s^2D_{g\to\gamma})$ contributions. In fact, $\alpha_sD_{q\to\gamma}$ is still 
larger than $D_{g\to\gamma}$, so we will include in our plots the $\alpha_s$ corrections to 
the color octet contributions (\ref{octetf}) proportional to $D_{q\to\gamma}$, which have been calculated in \cite{Maltoni:1998nh}.
In addition, the coefficients for the octet $\phantom{}^3P_J$ contributions have large numerical factors, causing these terms to be more important than 
the color singlet contributions. 
Let us finally notice that the $\alpha_s$ corrections to the singlet rate  will produce terms of $\mathcal{O}(\alpha_s^4D_{q\to\gamma})$, 
which from the considerations above are expected to be as important as  
the octet $\phantom{}^3S_1$ contribution. These $\alpha_s$ corrections to the singlet rate are unknown, which results in a large theoretical uncertainty in the fragmentation contributions.

For the quark fragmentation function we will use the LEP measurement \cite{Buskulic:1995au}  
\begin{equation}
D_{q\gamma}(z,\mu) = \frac{e_q^2\alpha(\mu)}{2\pi}
\left[P_{q\gamma}(z) \ln\left(\frac{\mu^2}{\mu_0^2(1-z)^2}\right) + C\right]
\end{equation}
where
\begin{equation}
C = -1-\ln(\frac{M_Z^2}{2\mu_0^2})\quad ;\quad P_{q\gamma}(z) = \frac{1+ (1-z)^2}{z}\quad
;\quad\mu_0=0.14^{+0.43}_{-0.12} {\rm\ GeV}
\end{equation} 
and for the gluon fragmentation function the model \cite{Owens:1986mp}. These are the same choices as in
\cite{Fleming:2002sr}. However, for the $O_8 (^1 S_0)$ and $O_8 (^3 P_0)$ matrix elements we will use our estimates in
\cite{GarciaiTormo:2004jw}
\begin{eqnarray}
\left.\left< \Upsilon (1S) \vert \mathcal{O}_8 (^1 S_0) \vert \Upsilon (1S) \right>\right|_{\mu=M} & \sim & 0.004\,GeV^3\\
\left.\left< \Upsilon (1S) \vert \mathcal{O}_8 (^3 P_0) \vert \Upsilon (1S) \right>\right|_{\mu=M} & \sim & 0.08\,GeV^5
\end{eqnarray}

The above numbers are obtained in an $MS$ scheme from dimensionally regularized US loops 
only. The value we assign to the $S$-wave matrix element is compatible with the recent 
(quenched) lattice determination (hybrid algorithm)\cite{Bodwin:2005gg}. Notice that we do not assume that a suitable 
combination of these matrix elements is small, as it was done in \cite{Fleming:2002sr}. 
The $O_8 (^3 S_1)$ matrix element can be extracted from a lattice determination of the 
reference above \cite{Bodwin:2005gg}. Using the wave function at the origin with the $\als^2 (\mu_s )$ corrections included, we obtain, 
\be
\left.\left< \Upsilon (1S) \vert \mathcal{O}_8 (^3 S_1) \vert \Upsilon (1S)
  \right>\right|_{\mu=M}  \sim 0.00026\,GeV^3 
\label{lattice}
\ee
which differs from the estimate using NRQCD $v$ scaling by more than two orders of magnitude:
\begin{equation}
\left.\left< \Upsilon (1S) \vert \mathcal{O}_8 (^3 S_1) \vert \Upsilon (1S) \right>\right|_{\mu=M}\sim
v^4\left.\left< \Upsilon (1S) \vert \mathcal{O}_1 (^3 S_1) \vert \Upsilon (1S) \right>\right|_{\mu=M}\sim 0.02\,GeV^3
\label{vscaling}
\end{equation}
(we have taken $v^2\sim0.08$), which was used in ref. \cite{Fleming:2002sr}. The description of 
data turns out to be better with the estimate (\ref{vscaling}). However, this is
not very significant, since, as mentioned before, unknown NLO contributions are expected to
be sizable.   
  
In the $z\rightarrow 0$ region soft radiation becomes dominant and the fragmentation
contributions completely dominate the spectrum in contrast with the direct contributions \cite{Catani:1994iz}. Note
that, since the fragmentation contributions have an associated bremsstrahlung spectrum, they can not be safely
integrated down to $z=0$; that is $\int_0^1dz\frac{d\Gamma^{frag}}{dz}$ is not an infrared safe observable. In any case
we are not interested in regularizing such divergence because the resolution of the detector works as a physical
cut-off.

\section{Scale setting and error analysis}\label{secphen}

Formula (\ref{mergingNLO}) requires $d\Gamma^{e}/dz$ for all values of $z$. The color octet shape functions, however, were calculated in the end-point region under the assumption that $M\sqrt{1-z}\sim \gamma$, and the scale of the $\als$ was set accordingly. When $z$ approaches the central region  
$M\sqrt{1-z}\gg \gamma$, and hence some $\als$ will depend on the scale $M\sqrt{1-z}$ and others on $\gamma$ (we leave aside the global $\als (\mu_u)$, which will be discussed below). In order to decide the scale we set for each $\als$ let us have a closer look at the formula (\ref{expando}). We see that all terms have a common factor $\gamma^3$. This indicates that one should extract $\gamma^3$ factors in the shape functions, the $\als$ of which should stay at the scale $\mu_s$. This is achieved by extracting $\gamma^{3/2}$ in $I_S$ and $I_P$. If we set the remaining $\als$ to the scale 
$\mu_p=\sqrt{m(M(1-z)/2-E_1)}$,
we will reproduce (\ref{expando}) when approaching to the central region, except for the relativistic correction, the $\als$ of which will be at the scale $\mu_p$ instead of at the right scale $\mu_s$. We correct for this by making
the following substitution
\be
S_{P1}\longrightarrow S_{P1}+{\als (\mu_u)\over 6\pi N_c}{\gamma^3\over\pi}\left(\log {k_+^2\over \mu_c^2}-1\right)\left( {4\gamma^2\over 3m}- {m C_f^2\als^2 (\mu_p)\over 3} \right)
\ee
Notice that the replacements above are irrelevant as far as the end-point region is concerned, but important for the shape functions to actually (numerically) approach the expressions (\ref{expando}) in the central region, as they should.

We use the following values of the masses for the plots: $m=4.81$ GeV and $M=9.46$ GeV.
The hard scale $\mu_h$ is set to $\mu_h=M$. The soft scale $\mu_s = m C_f\alpha_s$ is to be used for the $\alpha_s$
participating in the bound state dynamics, we have $\alpha_s(\mu_s)=0.28$. The US scale $\mu_u$, arising from
the couplings of the US gluons, is set to $\mu_u=\sqrt{2M(1-z)\left(\frac{M}{2}(1-z)-E_1\right)}$ (as discussed
in section \ref{secdirect}). We have used the \verb|Mathematica| package \verb|RunDec| \cite{Chetyrkin:2000yt} to obtain
the (one loop) values of $\alpha_s$ at the different scales.

\begin{figure}
\centering
\includegraphics[width=15cm]{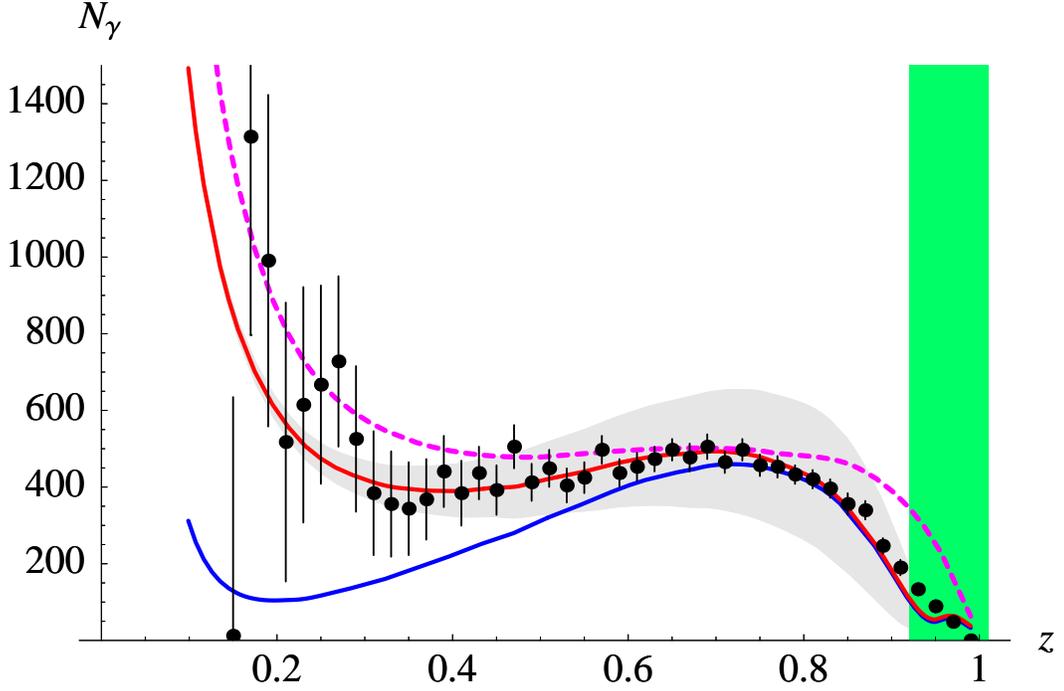}
\caption{Photon spectrum. The points are the CLEO data
  \cite{Nemati:1996xy}. The solid lines are the NLO merging in section
  \ref{subsubsecNLO} plus the fragmentation contributions: the red (light)
  line and the blue (dark) line are obtained by using (\ref{vscaling}) and (\ref{lattice}) for $\left< \Upsilon (1S) \vert
  O_8 (^3 S_1) \vert \Upsilon (1S) \right>$  respectively.
The grey shaded region is obtained by varying
  $\mu_{c}$ by $\sqrt{2^{\pm 1}}\mu_{c}$. The green shaded region on the right
shows the zone where the calculation of the shape functions is not
  reliable (see section \ref{subsubsecCO}). The pink dashed line is the result
  in \cite{Fleming:2002sr}, where only color singlet contributions were included in the direct contributions.}\label{total}
\end{figure}

Our final plot in Fig. \ref{total} is obtained by using the merging formula (\ref{mergingNLO}) at NLO with the $\als^2 (\mu_s)$  corrections to the wave function at the origin included for the direct contributions plus the fragmentation contributions in section \ref{secfrag} 
including the first $\als$ corrections in $C_q$ and
using the estimate (\ref{vscaling}) for the $\left< \Upsilon (1S) \vert
  \mathcal{O}_8 (^3 S_1) \vert \Upsilon (1S) \right>$ matrix element. The
error band is obtained by replacing $\mu_{c}$ by $\sqrt{2^{\pm 1}}\mu_{c}$. Errors associated to the large $\als^2 (\mu_s)$ corrections to the wave function at the origin, to possible large NLO color singlet contributions in the end-point region and to the fragmentation contributions are difficult to estimate and not displayed (see the corresponding sections in the text for discussions). The remaining error sources are negligible. As
usual experimental efficiency has been taken into account and the overall
normalization is a free parameter.

\section{Conclusions}\label{secconcl}

We have analyzed the photon spectrum in radiative $\Upsilon (1S)$ decays within an Effective Field Theory framework. For the direct contributions, the merging of the results for the central and upper end-point regions has been discussed in detail. We have shown how to consistently combine the complete LO results for the upper end-point region with the complete NLO ones for the central region. 
We have seen that the large $\als^2 (\mu_s)$ corrections to the wave function at the origin are important in order to get a good description of data. Otherwise,
parametrically subleading large contributions in the end-point region would be necessary. We would like to emphasize that our final results for the direct contributions are essentially parameter free: only the mass of the bottom quark $m$, the strong coupling constant $\als$, and the proper choice of subtraction scales (which appear in logarithms) are used as an input. 
For the fragmentation contributions, we have pointed out that if the commonly used model for the gluon fragmentation into a photon is appropriated, $\als$ corrections to the LO color singlet matrix element giving rise to a light quark which fragments into a photon may be as important as the LO results. Hence, fragmentation contributions suffer from large theoretical uncertainties. Nevertheless, if we put together the available theoretical results for these contributions with the ones for the direct contributions, an excellent description of data is achieved for the whole part of the spectrum where experimental errors are reasonable small.
Clearly, our results indicate that the introduction of a finite gluon mass \cite{Field:2001iu} is unnecessary.
One should keep in mind, however, that in order to have the theoretical errors under control higher order calculations are necessary both in the direct (end-point) and fragmentation contributions.

Before closing, let us mention that the inclusion of color octet contributions in the end-point
region together with the merging with the central region expression described in this work may
be useful for production processes like inclusive $J/\psi$ production in $e^+e^-$ machines
\cite{Fleming:2003gt,Lin:2004eu,Hagiwara:2004pf}.  

\begin{acknowledgments} We are grateful to Dave Besson and to Michael Kramer for providing us 
with the data of ref. \cite{Nemati:1996xy} and the numerical results of ref. 
\cite{Kramer:1999bf} respectively. Thanks are also given to Antonio Pineda for important 
discussions, and to Antonio Vairo for asking the right questions. We acknowledge financial support from a CICYT-INFN 2004 collaboration contract, Acciones Integradas HI2003-0362 (Spain-Italy),
the MCyT and Feder (Spain) grant FPA2001-3598, the
CIRIT (Catalonia) grant 2001SGR-00065 and the network EURIDICE (EU)
HPRN-CT2002-00311. X.G.T. thanks the Institute for Nuclear Theory at the
University of Washington for its hospitality and the Department of Energy for
partial support during the completion of this work. X.G.T. also acknowledges financial support from the Departament d'Universitats, Recerca i
Societat de la Informaci\'o of the Generalitat de Catalunya and the Fons Social Europeu.

\end{acknowledgments}

\appendix

\section{Definitions}\label{appdef}
In this appendix we collect the definitions for the formulas that appear in the paper.
\[
C_A=N_c\quad C_f=\frac{N_c^2-1}{2N_c}\quad\beta_0=\frac{\left(11C_A-2n_f\right)}{3}\quad\beta_1 = \frac{34C_A^2-10C_A n_f-6C_f n_f}{3}
\]
$N_c=3$ is the number of colors, $n_f=4$ is the number of light flavors.

\subsection{Definitions for formula (\ref{singres})}
\begin{equation}
f_{5/2}^{(n)} = \frac{n(n+1)(n+2)(n+3)}{9(n+3/2)}\quad;\quad f_{3/2}^{(n)} = \frac{(n+1)(n+2)}{n+3/2}
\end{equation}
\begin{equation}
r(\mu) = \frac{\alpha_s(\mu)}{\alpha_s(2m)}
\end{equation}
\begin{equation}
\gamma_\pm^{(n)} = \frac{\gamma_{gg}^{(n)} - \lambda^{(n)}_\mp}{\Delta}\quad;\quad \lambda^{(n)}_\pm = \frac{1}{2} \big[
\gamma^{(n)}_{gg} +  \gamma^{(n)}_{q\bar{q}} 
\pm \Delta \big]\quad;\quad\Delta = 
\sqrt{ (\gamma^{(n)}_{gg} -  \gamma^{(n)}_{q\bar{q}})^2 + 
                 4 \gamma^{(n)}_{gq} \gamma^{(n)}_{qg} }
\end{equation}
\begin{eqnarray}
\gamma^{(n)}_{q\bar{q}} &=& C_f \bigg[ \frac{1}{(n+1)(n+2)} -\frac{1}{2} - 2 \sum^{n+1}_{i=2} \frac{1}{i}
\bigg]\,\nonumber
\\
\gamma^{(n)}_{gq} &=&
\frac{1}{3}C_f \frac{n^2 + 3n +4}{(n+1)(n+2)}\,
\nonumber \\
\gamma^{(n)}_{qg} &=& 3 n_f \frac{n^2 + 3n +4}{n(n+1)(n+2)(n+3)}\,
\nonumber \\
 \gamma^{(n)}_{gg} &=&
 C_A \bigg[ \frac{2}{n(n+1)} + \frac{2}{(n+2)(n+3)}- \frac{1}{6} - 
 2 \sum^{n+1}_{i = 2} \frac{1}{i} \bigg] - \frac{1}{3}n_f
 \end{eqnarray}

\subsection{Definitions for formula (\ref{ocres})}
\begin{equation}
\ell\approx-\log(x-z)
\end{equation}
\begin{eqnarray}
g_1(\chi) &=& 
-\frac{2 \Gamma^{\rm adj}_1}{\beta_0\chi}\left[(1-2\chi)\log(1-2\chi) -2(1-\chi)\log(1-\chi)\right] \nonumber \\
g_2(\chi) &=& -\frac{8 \Gamma^{\rm adj}_2}{\beta_0^2}
  \left[-\log(1-2\chi)+2\log(1-\chi)\right] \nonumber\\
 && - \frac{2\Gamma^{\rm adj}_1\beta_1}{\beta_0^3}
   \left[\log(1-2\chi)-2\log(1-\chi)
  +\frac12\log^2(1-2\chi)-\log^2(1-\chi)\right] \nonumber\\
 &&+\frac{4\gamma_1}{\beta_0} \log(1-\chi) + 
 \frac{2B_1}{\beta_0} \log(1-2\chi) -\frac{4\Gamma^{\rm adj}_1}{\beta_0}\log n_0
 \left[\log(1-2\chi)-\log(1-\chi)\right]
\end{eqnarray}
\begin{equation}
\Gamma^{\rm adj}_1 =  C_A \quad;\quad 
\Gamma^{\rm adj}_2 =  
   C_A \left[ C_A \left( \frac{67}{36} - \frac{\pi^2}{12} \right) 
    - \frac{5n_f}{18} \right]  \quad;\quad 
B_1 = -C_A\quad;\quad \gamma_1 = -\frac{\beta_0}{4}\quad;\quad n_0=e^{-\gamma_E}
\end{equation}
\subsection{Definitions for formula (\ref{sp})}
\begin{equation}
I_{S}({k_+\over 2} +x)=m\sqrt{\gamma\over \pi}{\als N_c \over 2}{1\over 1-z'}\left( 1-{2z'\over 1+z'} \;\phantom{}_2F_1\left(-\frac{\lambda}{z'},1,1-\frac{\lambda}{z'},\frac{1-z'}{1+z'}\right)\right)
\end{equation}
\begin{displaymath}
I_{P}({k_+\over 2} +x)=\sqrt{\frac{\gamma^3}{\pi}}
{8\over 3}\left(
2-\lambda \right)\!\!\frac{1}{4(1+z')^3}\Bigg( 2(1+z')(2+z')+(5+3z')(-1+\lambda)+2(-1+\lambda)^2+
\end{displaymath}
\begin{equation}
\left.+\frac{1}{(1-z')^2}\left(4z'(1+z')(z'^2-\lambda^2)\left(\!\!-1+\frac{\lambda(1-z')}{(1+z')(z'-\lambda)}+\phantom{}_2F_1\left(-\frac{\lambda}{z'},1,1-\frac{\lambda}{z'},\frac{1-z'}{1+z'}\right)\right)\right)\right)
\end{equation}
\begin{equation}
S_{S}(k_+)={4\als (\mu_u)\over 3 \pi N_c} \left({ c_F\over 2m}\right)^2
\int_0^{\infty} dx \left( 2 \psi_{10}( {\bf 0})I_{S}({k_+\over 2} +x)- I_{S}^2({k_+\over 2} +x) \right)
\label{ss}
\end{equation}
\begin{equation}
S_{P1}(k_+)=
{\als (\mu_u)\over 6 \pi N_c}
\int_0^{\infty}\!\!\!dx\left( 2\psi_{10}( {\bf 0})I_P(\frac{k_+}{2}+x)-I_P^2(\frac{k_+}{2}+x) \right)
\label{sp1}
\end{equation}
\begin{equation}\label{ImTp}
S_{P2}(k_+)=
{\als (\mu_u)\over 6 \pi N_c}
\int_0^{\infty}\!\!\!dx \frac{8k_+x}{\left(k_++2x\right)^2}\left(
\psi^2_{10}( {\bf 0})-2\psi_{10}( {\bf 0})I_P(\frac{k_+}{2}+x)+I_P^2(\frac{k_+}{2}+x)\right)
\end{equation}
\begin{equation}
\gamma=\frac{mC_f\als}{2}\quad z'=\frac{\kappa}{\gamma}\quad-\frac{\kappa^2}{m}=E_1-\frac{k_+}{2}-x\quad\lambda=-\frac{1}{2N_cC_f} \quad E_1=-{\gamma^2\over m}
\label{defg}
\end{equation}
$c_F$ is the hard matching coefficient of the chromomagnetic interaction in NRQCD, it will be taken to 1. The renormalized expressions (see Appendix \ref{appregren}) in an $MS$ scheme read
\bea
S_{S}^{MS}(k_+)&=&{4\als (\mu_u)\over 3 \pi N_c} \left({ c_F\over 2m}\right)^2
\Bigg\{2 \psi_{10}( {\bf 0})\left(m\sqrt{\frac{\gamma}{\pi}}\frac{\alpha_sN_c}{2}\right)\Bigg(\int_0^{\infty}  \!\!\!\left(\widetilde{I}_{S}({k_+\over 2}
    +x)-\frac{1}{z'}-\left(-1+2\lambda\ln
      2\right)\frac{1}{z'^2}\right)dx-\nonumber\\
& & \left.-2\frac{\gamma}{\sqrt{m}}\sqrt{\frac{k_+}{2}+\frac{\gamma^2}{m}}
\right)-\left(m\sqrt{\frac{\gamma}{\pi}}\frac{\alpha_sN_c}{2}\right)^2\left(\int_0^{\infty}\!\!\!\left(\widetilde{I}_{S}^2({k_+\over 2}
        +x)-\frac{1}{z'^2}\right)dx
\right)\Bigg\}+
  \nonumber\\ & & +{c_F^2\als (\mu_u )\gamma^3 C_f^2 \als^2 (\mu_p)\over 3\pi^2 N_c m}(1-\lambda)\left( -2+\lambda (2\ln2 + 1)\right)\left(\ln\left(\frac{\mu_{pc}^2}{m\left(\frac{k_+}{2}+\frac{\gamma^2}{m}\right)}\right)\right)\\
S_{P1}^{MS}(k_+)&=&{\als (\mu_u)\over 6 \pi N_c} 
\Bigg\{2 \psi_{10}( {\bf 0})\left(\sqrt{\frac{\gamma^3}{\pi}}\frac{8}{3}(2-\lambda)\right)\Bigg(\int_0^{\infty}  \!\!\!\left(\widetilde{I}_{P}({k_+\over 2}
    +x)-\frac{1}{2z'}-\left(-\frac{3}{4}+\lambda\ln
      2-\frac{\lambda}{4}\right)\frac{1}{z'^2}\right)dx-\nonumber\\
& & \left.-\frac{\gamma}{\sqrt{m}}\sqrt{\frac{k_+}{2}+\frac{\gamma^2}{m}}
\right)-\left(\sqrt{\frac{\gamma^3}{\pi}}\frac{8}{3}(2-\lambda)\right)^2\left(\int_0^{\infty}\!\!\!\left(\widetilde{I}_{P}^2({k_+\over 2}
        +x)-\frac{1}{4z'^2}\right)dx\right)\Bigg\}
  +\nonumber\\ & & +{\als (\mu_u )\gamma^3 m C_f^2 \als^2 (\mu_p)\over 9\pi^2 N_c }
\left(\!\!-\frac{31}{6}+\lambda (4\ln2+\frac{19}{6})-\lambda^2 (2\ln 2 + {1\over 6})\right)\ln\!\!\left(\frac{\mu_{pc}^2}{m\left(\frac{k_+}{2}+\frac{\gamma^2}{m}\right)}\right)
+\nonumber\\ && + {2\als (\mu_u )\gamma^5  \over 9\pi^2 N_c m}\left(-\ln\left(\frac{\mu_c^2}{k_+^2}\right)\right)
\\
S_{P2}^{MS}(k_+)&=&{\als (\mu_u)\over 6 \pi N_c} 
\Bigg\{\psi^2_{10}( {\bf 0})k_+\left(-2+2\ln\left(\frac{\mu_c^2}{k_+^2}\right)\right)+\nonumber\\ & & +\int_0^{\infty}\!\!\!dx \frac{8k_+x}{\left(k_++2x\right)^2}\left(-2\psi_{10}( {\bf 0})I_P(\frac{k_+}{2}+x)+I_P^2(\frac{k_+}{2}+x)\right)\Bigg\}
\eea
and in the scheme where additional subtractions are carried out
\bea
S_{S}^{sub}(k_+)&=& S_{S}^{MS}(k_+)+{4\als (\mu_u)\over 3 \pi N_c} \left({ c_F\over 2m}\right)^2
2 \psi_{10}( {\bf 0})\left(m\sqrt{\frac{\gamma}{\pi}}\frac{\alpha_sN_c}{2}\right)2\frac{\gamma}{\sqrt{m}}\sqrt{\frac{k_+}{2}}\\
S_{P1}^{sub}(k_+)&=& S_{P1}^{MS}(k_+)+{\als (\mu_u)\over 6 \pi N_c} 
2 \psi_{10}( {\bf 0})\left(\sqrt{\frac{\gamma^3}{\pi}}\frac{8}{3}(2-\lambda)\right)\frac{\gamma}{\sqrt{m}}\sqrt{\frac{k_+}{2}}\\
S_{P2}^{sub}(k_+)&=&S_{P2}^{MS}(k_+)
\eea
where
\begin{eqnarray}
\widetilde{I}_S(\frac{k_+}{2}+x) & := &
\left(m\sqrt{\frac{\gamma}{\pi}}\frac{\alpha_sN_c}{2}\right)^{-1}{I}_S(\frac{k_+}{2}+x)\nonumber\\
\widetilde{I}_P(\frac{k_+}{2}+x) & := & \left(\sqrt{\frac{\gamma^3}{\pi}}\frac{8}{3}(2-\lambda)\right)^{-1}{I}_P(\frac{k_+}{2}+x)
\end{eqnarray}

\section{Regularization and renormalization}\label{appregren}

The shape functions (\ref{ss})-(\ref{ImTp}) are ultraviolet (UV) divergent and require regularization 
and renormalization. In ref. \cite{GarciaiTormo:2004jw} it was pointed out that using 
dimensional regularization (DR) for the US loop only was enough to regulate them. In fact, 
the expressions (\ref{ss})-(\ref{sp1}) implicitly assume that DR is used, otherwise linearly 
divergent terms proportional to $\psi^2_{10}({\bf 0}) $ would appear (which make (\ref{ss})-(\ref{sp1}) 
formally positive definite quantities). In addition an $MS$ scheme was used to subtract the 
poles. Since it turns out that the final outcome strongly depends on the details of this subtraction, let us spell out the procedure carried out in ref. \cite{GarciaiTormo:2004jw}. 
In order to isolate the $1/\varepsilon$ poles,  $I_S$ and $I_P$ were expanded
up to $\mathcal{O}(1/{z'}^2)$ . The result was subtracted and added to the integrand of
(\ref{ss})-(\ref{sp1}) (for (\ref{ImTp}) this is not necessary since the only
divergent piece is independent of $I_P$). The subtracted part makes the shape
functions finite. The added part contains linear and logarithmic UV
divergencies. The $1/\varepsilon$ ($D=4-2\epsilon$) poles displayed in
formulas (16) of ref. \cite{GarciaiTormo:2004jw}, and eventually subtracted,
were obtained by making $dx\rightarrow dx (x/ \mu)^{-\varepsilon}$ in
(\ref{ss})-(\ref{ImTp}). This was motivated by the fact that $x \sim {\bf
  k}_\perp^2$ (${\bf k}_\perp$ being the transverse momentum of the US gluon)
but differs from a standard $MS$ scheme. Linear divergences are set to zero as
usual in DR.

We have used here a regularization and renormalization scheme which is closer to the standard one in pNRQCD calculations. We have regulated both the US loop and the potential loops (entering in the bound state dynamics) in DR. We have identified US divergencies by taking the limit $D\rightarrow 4$ in the US loops while leaving the potential loops in $D$ dimensions \cite{Pineda:1997ie}. Potential divergencies are identified by taken $D\rightarrow 4$ in the potential loops once the US divergencies have been subtracted. It turns out that all divergencies in $S_{P2}$ are US and all divergencies in $S_S$ are potential. $S_{P1}$ contains both US and potential divergencies. The potential divergences related with the bound state dynamics can be isolated using the methods of ref. \cite{Czarnecki:1999mw}. The formulas corresponding to (16) of ref. \cite{GarciaiTormo:2004jw} in this scheme read
\begin{displaymath}
\left.S_{S}(k_+)\right\vert_{\varepsilon\rightarrow 0}\simeq
{c_F^2\als (\mu_u )\gamma^3 C_f^2 \als^2 (\mu_p)\over 3\pi^2 N_c m}(1-\lambda)\left( -2+\lambda (2\ln2 + 1)\right)\left(\frac{1}{\varepsilon}+\ln\left(\frac{\mu_{pc}^2}{m\left(\frac{k_+}{2}+\frac{\gamma^2}{m}\right)}\right)+\cdots\right)
\end{displaymath}
\bea
\left.S_{P1}(k_+)\right\vert_{\varepsilon\rightarrow 0} \!&\simeq &\!\!
{\als (\mu_u )\gamma^3 m C_f^2 \als^2 (\mu_p)\over 9\pi^2 N_c }
\left(\!\!-\frac{31}{6}+\lambda (4\ln2+\frac{19}{6})-\lambda^2 (2\ln 2 + {1\over 6})\right)\!\!\!\left(\frac{1}{2\varepsilon}+\ln\!\!\left(\frac{\mu_{pc}^2}{m\left(\frac{k_+}{2}+\frac{\gamma^2}{m}\right)}\right)
\!\!+\cdots\!\!
\right) \nonumber\\ && + {2\als (\mu_u )\gamma^5  \over 9\pi^2 N_c m}\left(-{1\over\varepsilon}-\ln\left(\frac{\mu_c^2}{k_+^2}\right)+\cdots\right)
\eea
\begin{equation}
\left.
S_{P2}(k_+)\right\vert_{\varepsilon\rightarrow 0}\simeq
{\als (\mu_u ) k_+ \gamma^3\over 3\pi^2 N_c}
\label{singular}
\left(\frac{1}{\varepsilon}
+\ln\left(\frac{\mu_c^2}{k_+^2}\right)+\cdots
\right)
\end{equation}
For simplicity, we have set $D=4$ everywhere except in the momentum integrals. 
$\mu_p$ is defined in section \ref{secphen}. $\mu_c$ and $\mu_{pc}$ are the subtraction points of the US and
potential divergencies respectively. If we subtract the $1/\varepsilon$ poles
and set $\mu_c=M\sqrt{1-z}$ and  $\mu_{pc}=\sqrt{m\mu_c}$ we obtain exactly
the same result as in ref. \cite{GarciaiTormo:2004jw} for what the potential
divergences is concerned\footnote{We assume that the correlation of scales
  advocated in \cite{Luke:1999kz} (see \cite{Pineda:2001et} for the
  implementation in our framework) must also be taken into account here.}. For the US divergences there is a factor
$\ln\left(\frac{\mu_c}{2k_+}\right)$ of difference with respect to the
previous scheme.

In ref. \cite{GarciaiTormo:2004kb} an additional subtraction
related to linear divergencies was made. This subtraction was necessary in
order to merge smoothly with the results in the central region. We will also
need this subtraction here when merging at LO, as discussed in subsection \ref{subsecmatch}. 
We use
\begin{displaymath}
\int_0^{\infty}\!\!\!\!\!\!dx\,
\frac{1}{z'}
\longrightarrow
-2\frac{\gamma}{\sqrt{m}}\left[\sqrt{\frac{k_+}{2}+\frac{\gamma^2}{m}}-\sqrt%
{\frac{k_+}{2}}\right]
\end{displaymath}
, which differ from the $MS$ scheme by the subtraction of the second term in the square brackets.

\section{The shape functions in the 
central region}\label{appcenreg}

When $z$ approaches the central region from the upper end-point, the shape functions should reduce to matrix elements of NRQCD operators multiplied by the corresponding matching coefficients. We will see here that this is indeed the case. 

Let us first consider the $S$-wave octet shape function as defined in \cite{GarciaiTormo:2004jw}
\begin{equation}\label{Swave}
I_{S}({k_+\over 2} +x):=\int d^3 {\bf x} \psi_{10}( {\bf x})\left( 1-{\frac{k_+}{2}+x \over h_o -E_1 +{k_+ \over 2}+x}\right)_{{\bf x},{\bf 0}}
\end{equation}
$h_o={\bf p}^2/m+V_o$, $V_o=\als/(2N_c\vert {\bf r}\vert )$.
When 
$z$ approaches the central region, $k_+\sim M(1-z) \gg -E_1$ and the larger three momentum scale is $M\sqrt{1-z}\gg\gamma$, the typical three momentum in the bound state. Therefore we can treat the Coulomb potential in (\ref{Swave}) as a perturbation when it is dominated by this scale. It is convenient to proceed in two steps. First we write $h_o=h_s + (V_o-V_s)$, where $h_s={\bf p}^2/m+V_s$, $V_s=-\als C_f/\vert {\bf r}\vert$,  and expand $V_o-V_s$. This allows to set $h_s -E_1$ to zero in the left-most propagator and makes explicit the cancellation between the first term in the series and the first term in (\ref{Swave}). It also makes explicit that the leading term will be proportional to $\als (M\sqrt{1-z})$. Second, we expand $V_s$ in $h_s={\bf p}^2/m+V_s$. In addition, since $M\sqrt{1-z} \gg\gamma$, the wave function can be expanded about the origin. Only the first term in both expansion is relevant in order to get (\ref{expando}). 

Consider next the $P$-wave shape functions as defined in \cite{GarciaiTormo:2004jw}
\begin{equation}\label{Pwave}
I_{P}({k_+\over 2} +x):=-\frac{1}{3}\int d^3 {\bf x} {\bf x}^i \psi_{10}( {\bf x})\left(\left(1- {\frac{k_+}{2}+x \over h_o -E_1 +{k_+ \over 2}+x} \right)\bfnabla^i \right)_{{\bf x},{\bf 0}}
\end{equation}
In order to proceed analogously to the $S$-wave case, we have first to move the ${\bf x}^i$ away from the wave function
\begin{displaymath}
I_{P}({k_+\over 2} +x)=\psi_{10}({\bf 0})+\frac{\frac{k_+}{2}+x}{3}\int d^3 {\bf x}\psi_{10}( {\bf x})\left\{\frac{1}{h_o -E_1 +{k_+ \over 2}+x}{\bf x} \bfnabla+\right.
\end{displaymath}
\begin{equation}\label{Pwave2}
\left. +{\frac{1}{h_o -E_1 +{k_+ \over 2}+x}}\left(-\frac{2\bfnabla^i}{m}\right){\frac{1}{h_o -E_1 +{k_+ \over 2}+x}}\bfnabla^i \right\}
\end{equation}
For the left-most propagators we can now proceed as before, namely expanding $V_o-V_s$. Note that the leading contribution in this expansion of the second term above exactly cancels against the first term. Of the remaining contributions of the second term only the next-to-leading one ($\mathcal{O}(\als)$) is relevant to obtain (\ref{expando}).
Consider next the leading order contribution in this expansion of the last term. It reads 
\begin{displaymath}
-\frac{2}{3m}\int d^3 {\bf x}\psi_{10}( {\bf x})\left\{\bfnabla^i \frac{1}{h_o -E_1 +{k_+ \over 2}+x}\bfnabla^i\right\}=
-\frac{2}{3m}\int d^3 {\bf x}\psi_{10}( {\bf x})\left\{\left( \frac{1}{h_o -E_1 +{k_+ \over 2}+x}\bfnabla^i-\right.\right.
\end{displaymath}
\begin{equation}\label{Pwave3}
\left.\left.-\frac{1}{h_o -E_1 +{k_+ \over 2}+x}\bfnabla^iV_o\frac{1}{h_o -E_1 +{k_+ \over 2}+x}\right)\bfnabla^i \right\}
\end{equation}
Now we proceed as before with the left-most propagators, namely expanding $V_o-V_s$. The leading order contribution of the first term above produces the relativistic correction $\mathcal{O}(v^2)$ of (\ref{expando}). The next-to-leading contribution of this term and the leading order one of the second term are $\mathcal{O}(\als)$ and also relevant to (\ref{expando}). The next-to-leading order contribution of the last term in (\ref{Pwave2}) in the $V_o-V_s$ expansion of the left-most propagator is also $\mathcal{O}(\als )$ and relevant to (\ref{expando}).

\end{document}